\documentclass[sn-mathphys,Numbered]{sn-jnl}

\usepackage{xcolor}
\usepackage{graphicx}
\usepackage{amsthm}
\usepackage{amsmath}
\usepackage{amssymb}
\usepackage[numbers,sort&compress]{natbib}
\usepackage[percent]{overpic}
\usepackage[T1]{fontenc}
\usepackage[utf8]{inputenc}
\usepackage{caption}
\usepackage{etoc}
\usepackage{pdfpages}

\theoremstyle{definition}
\newtheorem{definition}{Definition}[section]

\begin{document}

\title[Topological analysis of the permeability tensor]{Topological analysis of permeability tensor in anisotropic porous media}

\author*[1,2]{\fnm{Rafa{\l}} \sur{Topolnicki}}\email{rafal.topolnicki@uwr.edu.pl}
\author[1]{\fnm{Pawe{\l}} \sur{D{\l}otko}}
\author[3,4]{\fnm{Maciej} \sur{Matyka}}

\affil[1]{Dioscuri Center in Topological Data Analysis, Center of Trustworthy AI for Life Sciences, International Research Agendas Programme, Warsaw University, 00-927, Warsaw, Poland}
\affil[2]{Institute of Experimental Physics, Faculty of Physics and Astronomy, University of Wroc{\l}aw, pl.~M.~Borna 9, 50-204, Wroc{\l}aw, Poland}
\affil[3]{Institute of Theoretical Physics, Faculty of Physics and Astronomy, University of Wroc{\l}aw, pl.~M.~Borna 9, 50-204, Wroc{\l}aw, Poland}
\affil[4]{Parallel and Distributed Systems Laboratory, Jo\v{z}ef Stefan Institute, Jamova cesta 39, 1000, Ljubljana, Slovenia}

\abstract{                                                                               
The permeability of an anisotropic porous medium is a tensor whose off-diagonal components, which describe the deflection of the macroscopic flux away from the driving force, are small and change sign under reflections. We ask whether direction-aware topological data analysis can predict the full permeability tensor of three-dimensional porous media directly from their geometry. We introduce two direction-aware topological descriptors for flow problems: a cone-based filtration, adapted from our earlier work on the elastic stiffness tensor, and a throat-scale filtration built on the distance transform of the pore space. Using permeability tensors computed with the lattice Boltzmann method, we compare them with a comprehensive set of non-topological baselines, comprising statistical microstructure descriptors and physics-based permeability estimators. The topological descriptors are most informative for the off-diagonal components, which the baselines capture least well. There the throat-scale filtration alone outperforms every non-topological descriptor set, including the union of all statistical descriptors and physics-based estimators, and adding it to the baselines consistently improves the prediction. We also confirm the approach on real rock data.
}     


\maketitle

\section{Introduction}

The permeability of a porous medium, its ability to conduct fluid under an applied pressure gradient or body force, governs oil and gas recovery~\cite{Zhong21}, the feasibility of CO$_2$ storage~\cite{Cossins23}, groundwater flow~\cite{Bear1972} and filtration. It is a standard parameter in the characterization of natural and engineered porous media~\cite{Bear1972, Clavaud2008}.

Anisotropy turns permeability into a tensor. For creeping flow, the macroscopic flux $\mathbf{q}$ is related to the driving force by the tensorial form of Darcy's law~\cite{Bear1972, Koza2009},
\begin{equation}
\mathbf{q} = -\frac{1}{\mu}\,\mathbf{K}\left(\nabla P - \varrho\,\mathbf{g}\right),
\label{eq:darcy_tensor}
\end{equation}
where $\mu$ is the dynamic viscosity, $\varrho$ the fluid density, $\mathbf{g}$ the external body force per unit mass, and $\mathbf{K}$ the permeability tensor. For Stokes flow $\mathbf{K}$ is symmetric and positive definite~\cite{Bear1972}. In three dimensions it therefore has six independent components, the diagonal $k_{xx}, k_{yy}, k_{zz}$ and the off-diagonal $k_{xy}, k_{xz}, k_{yz}$. Whenever an off-diagonal component is non-zero, the flux and the force point in different directions: a fluid pushed along $x$ also flows along $y$. Koza, Matyka and Khalili~\cite{Koza2009} found this effect even in statistically isotropic samples of the sizes typically used in simulations, where it arises from the finite system size and decays as the system grows. Matyka \emph{et al.}~\cite{Matyka2013} extended the analysis to three-dimensional periodic models. In anisotropic media such as layered or sheared rocks, the effect is a property of the material itself~\cite{Clavaud2008}.

The off-diagonal components are small compared with the diagonal ones, and they vanish identically whenever the principal axes of the structure coincide with the coordinate axes. Under a reflection of the sample the off-diagonal components change sign, while the diagonal ones do not. A descriptor that does not change under reflection, such as porosity, the specific surface or the scalar Kozeny--Carman estimate~\cite{Bear1972, Dullien1992}, therefore cannot predict the off-diagonal components. Predicting them requires a descriptor that depends on direction, including its sign.

Machine-learning models trained directly on images of the structure have been shown to predict scalar permeability, tortuosity, velocity fields and elastic constants of porous media~\cite{Graczyk20, Topolnicki2026PINN, Zorkaltsev2026}. Such models give little insight into which geometric features control the prediction. Topological data analysis (TDA) quantifies the connectivity, loops and cavities of a structure as a function of scale, and it has been applied to permeability, trapping and transport in porous media~\cite{Robins2016, Herring2019, Moon2019, Thompson2023, Lisitsa2020}. Standard topological descriptors are, however, invariant under rotations and reflections. In our recent work~\cite{Topolnicki2026TDA} we removed this limitation by embedding the loading direction into the filtration functions for persistent homology (PH) and Euler characteristic profiles (ECP). These direction-aware descriptors were the only method that retained predictive power on the off-diagonal, normal--shear coupling terms of the elastic stiffness tensor. For flow the correspondence is exact, with the diagonal permeabilities in the role of the uniaxial stiffnesses and the off-diagonal ones in the role of the coupling terms, small and sign-carrying in both cases.

In this work we apply the direction-aware framework to the permeability tensor of three-dimensional porous structures. Every structure is a triply periodic $80\times80\times80$ voxel grid, and its tensor is computed with the lattice Boltzmann method (LBM). Our contributions are the following:
\begin{itemize}
\item[1)] We construct a dataset of 5000 periodic three-dimensional porous structures with controlled anisotropy in a random principal frame. The full permeability tensor of each structure comes from three independent solves with a D3Q19 LBM solver.
\item[2)] We introduce a throat-scale, direction-aware filtration (THR) built on the distance transform of the pore space and motivated by critical-path analysis~\cite{Katz1986}. We compare it with the cone-based filtration of our earlier work~\cite{Topolnicki2026TDA}, referred to as TDA.
\item[3)] We benchmark both topological descriptors against directional statistical descriptors (porosity profile, two-point correlation, fabric tensor) and against physics-based estimators (a tensorial Kozeny--Carman estimate and directional critical-path radii), alone and in combination. The comparison is paired and fold-by-fold, in $R^2$ and in the mean absolute error.
\end{itemize}

Section~\ref{sec:dataset} describes the lattice Boltzmann solver, how the dataset is generated and how the permeability tensor is computed. The non-topological baselines and the two direction-aware topological descriptors are defined in Section~\ref{sec:descriptors}, which also gives an introduction to persistent homology for readers not familiar with it, and the details of model training and evaluation are given in Section~\ref{sec:training}. Results follow in Section~\ref{sec:results}, and Section~\ref{sec:summary} summarizes the results and outlines the next steps.

\section{Dataset}
\label{sec:dataset}

\subsection{Pore-scale flow}

We describe porous samples at the pore scale, where the solid phase fills the space up to the desired porosity of the sample. We assume periodic boundary conditions in all three directions. The flow is driven by a constant body force, and with no-slip boundary conditions on the solid surface the system stabilizes to a stationary flow. The governing equations for the flow of an incompressible fluid in the pores are the Navier--Stokes equations,
\begin{equation}
\nabla \cdot \mathbf{u}=0, \qquad
\varrho(\mathbf{u} \cdot \nabla \mathbf{u})=-\nabla p+\mu \nabla^2 \mathbf{u}+\mathbf{f},
\label{eq:nse}
\end{equation}
where $\mathbf{u}$ is the velocity, $\mu$ the dynamic viscosity, $p$ the pressure, $\varrho$ the density, and $\mathbf{f}$ the external force density. We operate in the regime of low Reynolds numbers, in which the macroscopic flux satisfies Darcy's law, Eq.~\eqref{eq:darcy_tensor}.

\subsection{The Lattice Boltzmann Method}

To solve the flow problem we use the mesoscopic Lattice Boltzmann Method~\cite{Succi2001, Kruger2017}. Instead of solving the Navier--Stokes equations directly, the method evolves the particle distribution function $f_i(\mathbf{x},t)$ on a regular grid, where it represents the probability of finding particles with a given lattice velocity $\mathbf{e}_i$ at a given location and time. The method is popular in porous-media flows primarily due to its simplicity and its ability to handle complex geometries with local no-slip boundary conditions~\cite{Succi2001}. The distribution function is transported according to
\begin{equation}
f_i(\mathbf{x}+\mathbf{e}_i,t+\delta t)=f_i(\mathbf{x},t)+\frac{f_i^{eq}(\mathbf{x},t)-f_i(\mathbf{x},t)}{\tau},
\label{eq:lbm}
\end{equation}
where $\tau$ is the relaxation time and $f_i^{eq}$ is the distribution function in equilibrium. The collision term on the right-hand side is the single-relaxation-time Bhatnagar--Gross--Krook (BGK) approximation of the collision operator~\cite{Bhatnagar1954, Qian1992}. Its moments give the macroscopic density $\varrho(\mathbf{x},t)$ and velocity $\mathbf{u}(\mathbf{x},t)$ fields,
\begin{equation}
\varrho(\mathbf{x},t)=\sum_i f_i(\mathbf{x},t), \qquad
\varrho(\mathbf{x},t)\mathbf{u}(\mathbf{x},t)=\sum_i f_i(\mathbf{x},t)\mathbf{e}_i .
\label{eq:moments}
\end{equation}
The flow is driven by a constant body force, and no-slip conditions on the solid surface are imposed with the halfway bounce-back rule. All quantities are expressed in lattice units.

The solver is stopped when the domain-averaged flux has reached a stationary value. Every 2000 steps, each component of the flux $\mathbf{q}$ is compared with its value 2000 steps earlier, and the run is stopped once all three components have moved by less than $10^{-5}\,|\mathbf{q}|$ in two consecutive checks (and not before step 4000). The tolerance is relative to the norm of the flux rather than to the component tested, because the transverse components, which carry the off-diagonal permeabilities, are small by construction.

\subsection{Computing the permeability tensor}

The permeability tensor is computed following the procedure described in~\cite{Koza2009, Matyka2013}. For every structure three independent simulations are performed, with the body force applied along $+x$, $+y$ and $+z$, respectively. In the simulation with the force along a given axis $j\in\{x,y,z\}$, the macroscopic flux is the Darcy (superficial) velocity, that is, the velocity averaged over the whole domain, $q_i = \langle u_i\rangle$, and the $j$-th column of the tensor follows from Eq.~\eqref{eq:darcy_tensor},
\begin{equation}
K_{ij} = \frac{\mu\, q_i^{(j)}}{F}, \qquad i,j\in\{x,y,z\},
\label{eq:Kij}
\end{equation}
where $q_i^{(j)}$ is the $i$-th component of the flux in the simulation forced along $j$, and $F=|\mathbf{f}|$ is the magnitude of the body force density of Eq.~\eqref{eq:nse}. On the periodic domain there is no macroscopic pressure gradient, $\nabla P=0$, and the body force plays the role of $\varrho\,\mathbf{g}$ in Eq.~\eqref{eq:darcy_tensor}.
Because Stokes flow is reciprocal, the exact tensor is symmetric, $K_{ij}=K_{ji}$, and the two independent estimates of each off-diagonal entry provide a free error bar. We report the symmetrized off-diagonal components, $k_{ij} = (K_{ij}+K_{ji})/2$ for $i\neq j$, and write $k_{xx}\equiv K_{xx}$ etc.\ for the diagonal ones. The reciprocity residual, $|K_{ij}-K_{ji}|/\mathrm{tr}\,\mathbf{K}$, has a median of $1.7\times10^{-7}$ and a maximum of $9.4\times10^{-6}$ across the dataset, so the label noise on the off-diagonal components is negligible. As an additional check of the tensor assembly, the flux measured in a fourth simulation with the force along $(1,1,1)$ was reproduced by $\mathbf{K}$ to a relative error between $10^{-7}$ and $1.4\times10^{-6}$ (tested on the structures of Figure~\ref{fig:examples} and on a pilot structure). All permeabilities are reported in lattice units.

\subsection{Porous samples}

Three-dimensional porous structures were represented as binary voxel images with a resolution of $80\times80\times80$. As in our earlier works~\cite{Topolnicki2026PINN, Topolnicki2026TDA}, the structures were generated by superimposing standing waves with random wave vectors $\mathbf{q}_i$ and phases $\phi_i$:
\begin{equation}
f(\mathbf{x}) = \sqrt{\frac{2}{N}} \sum_{i=1}^{N} \cos(2\pi\,\mathbf{q}_i \cdot \mathbf{x} + \phi_i), \quad \mathbf{x} \in [0,1)^3,
\label{eq:random_trig}
\end{equation}
where the phases $\phi_i$ are sampled uniformly from $[0,\pi]$ and the number of modes $N$ is drawn uniformly from $[20,80]$. The wave vectors are first drawn isotropically, with uniformly distributed directions and magnitudes uniform in $[k_0-1,\,k_0+1]$ (in cycles per domain), where the central frequency $k_0$ is drawn uniformly from $[3,7]$ for each structure. Anisotropy is introduced in Fourier space by a volume-preserving ellipsoidal map,
\begin{equation}
\mathbf{q}_i \;\to\; \mathbf{R}\,\mathrm{diag}(s_1,s_2,s_3)\,\mathbf{R}^{T}\mathbf{q}_i,
\qquad (s_1,s_2,s_3) = (AB)^{1/3}\left(\frac{1}{A},\frac{1}{B},1\right),
\label{eq:squash}
\end{equation}
after which the components are rounded to integers, which makes the field, and hence the structure, exactly periodic. Here $\mathbf{R}$ is a uniformly distributed random rotation, drawn independently for every structure, so that the principal axes of the structure are not aligned with the grid. The wavelength is the longest along the first principal axis, so solid features elongate along it; $A\geq1$ is the elongation ratio of the first to the third principal axis and $B\in[1,A]$ that of the second to the third, interpolating between rod-like ($B=1$) and plate-like ($B=A$) anisotropy. The dataset consists of three equally sized classes: isotropic ($A=B=1$, 1667 structures), moderately anisotropic ($A\in[1.5,2.5]$, 1667 structures), and strongly anisotropic ($A\in[2.5,4.0]$, 1666 structures). A two-dimensional illustration of the map, showing how it acts on the wave vectors and how the same structure elongates as $A$ grows, is given in Section~S5 of the Supplementary Information.

Given the random function~\eqref{eq:random_trig}, a point $\mathbf{x}$ belongs to the solid phase $\mathcal{B}$ if $f(\mathbf{x})>\varepsilon$ and to the pore phase $\mathcal{P}$ if $f(\mathbf{x})\le\varepsilon$, where the threshold $\varepsilon$ is chosen to achieve the desired porosity $\varphi$, sampled uniformly from $0.65\le\varphi\le0.90$. The porosity is defined as the ratio of pore voxels to the total number of voxels. Structures in which the pore space does not percolate in all three directions (tested with face connectivity and periodic boundary conditions) were rejected and regenerated; in practice, no structure of the final dataset required a second attempt. A total of 5000 structures were generated, and for all of them the three LBM solves converged.

\begin{figure}
\centering
\begin{minipage}{0.48\linewidth}\centering
\includegraphics[width=\linewidth]{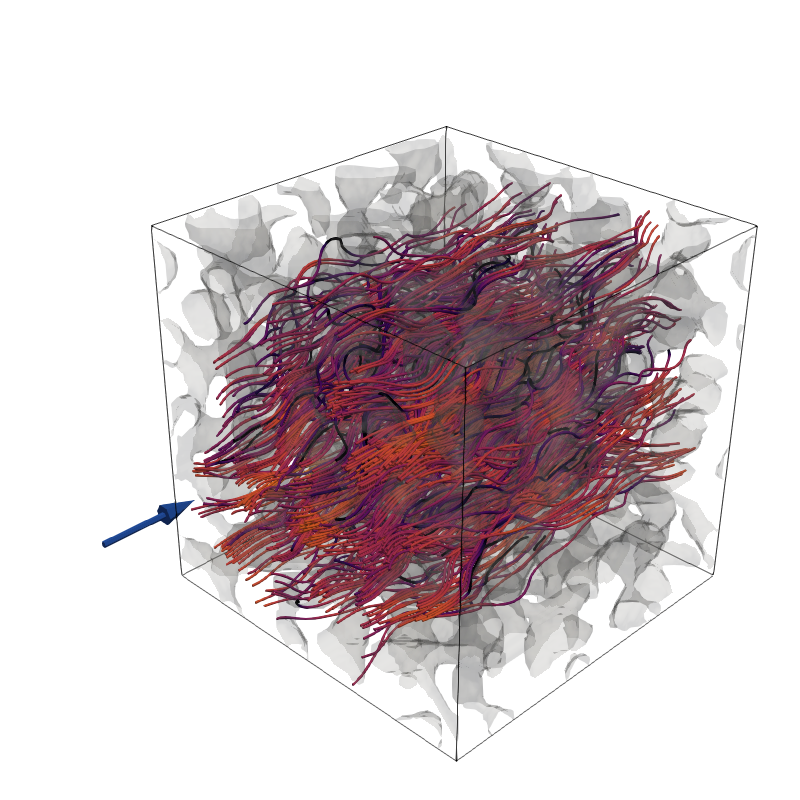}\\ a)
\end{minipage}\hfill
\begin{minipage}{0.48\linewidth}\centering
\includegraphics[width=\linewidth]{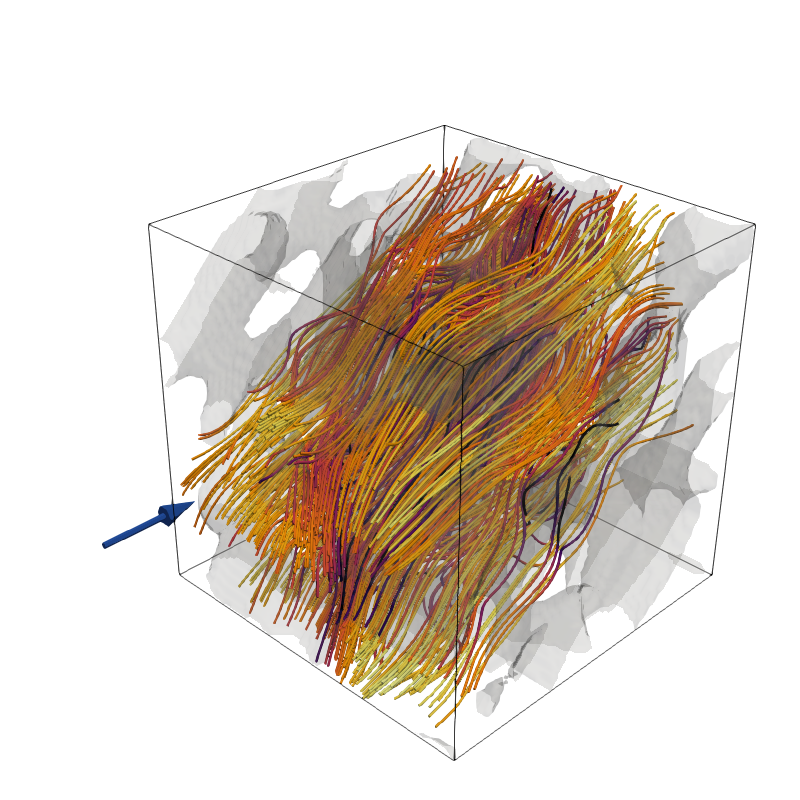}\\ b)
\end{minipage}
\caption{Two exemplary structures of the dataset, both at porosity $0.78$, with streamlines of the LBM velocity field, coloured by speed (logarithmic scale, the same for both panels); the blue arrow shows the direction of the body force, along $x$ in (a) and along $y$ in (b). In (b) the streamlines turn towards $z$, which is the large $k_{yz}$ seen directly. (a) An isotropic structure ($A=B=1$); its off-diagonal permeabilities are below $1.4\%$ of the mean permeability and are a pure finite-size effect~\cite{Koza2009,Matyka2013}. (b) A strongly anisotropic structure ($A=2.97$, $B=1.76$) with $k_{yz}=4.91$ lattice units, about half of its diagonal components; the direction of its largest principal permeability lies close to the $[011]$ diagonal.}
\label{fig:examples}
\end{figure}

Two exemplary structures with the corresponding LBM velocity fields are shown in Figure~\ref{fig:examples}.

Section~S1 of the Supplementary Information describes the dataset in more detail: it gives the distributions of the six components of the permeability tensor in the three anisotropy classes and the scales at which they occur. The diagonal components range from $0.5$ to $43$ lattice units, while the off-diagonal components are distributed symmetrically around zero with a standard deviation of about $0.7$ lattice units.

\section{Descriptors of the structures}
\label{sec:descriptors}

All directional descriptors were computed for the same set of nine integer lattice directions $\hat d = [hkl]$, written for the (unnormalized) direction vector $(h,k,l)$: $[100]$, $[010]$, $[001]$, $[110]$, $[1\bar10]$, $[101]$, $[10\bar1]$, $[011]$ and $[01\bar1]$, namely the three Cartesian axes and the six \emph{signed} face diagonals. An overbar denotes a negative coordinate: \([hk\bar{l}]\) represents the direction \((h,k,-l)\). This set differs from the seven directions used in~\cite{Topolnicki2026TDA}: the body diagonal is dropped and the face diagonals are taken in pairs of opposite sign. The pair is essential for the off-diagonal components: $k_{xy}$ measures the contrast between the conductances along $[110]$ and $[1\bar10]$, and a descriptor that only sees one of the two diagonals cannot express the sign of $k_{xy}$. For each structure the per-direction descriptors are concatenated into a single feature vector.

\subsection{Non-topological baselines}

To determine whether the predictive power of the topological descriptors arises from their topological content or merely from directional information in any form, we use two families of non-topological baselines, statistical microstructure descriptors and physics-based estimators. For a fair comparison, all of them are computed in the same way as the topological descriptors below: every descriptor that depends on direction is evaluated separately for each of the nine directions $\hat d$, and all quantities are computed with periodic boundary conditions. Their full definitions are given in Section~S2 of the Supplementary Information.

\paragraph{Statistical microstructure descriptors.} As in~\cite{Topolnicki2026TDA}, we use the porosity profile along $\hat d$ (Porosity)~\cite{Torquato2002}, the two-point correlation function along $\hat d$ (TPC)~\cite{Torquato2002, PhysRevE.76.031110}, which measures the characteristic feature size in that direction, and the fabric tensor of the interface normals together with its directional profile (Fabric tensor, Fabric profile)~\cite{Kanatani1984, Cowin1985}, which describe the orientation of the solid--void interface. Their union ($1819$ features) is denoted \emph{Stats}.

\paragraph{Physics-based estimators.} We also use two established physical estimators of permeability: the Kozeny--Carman estimate (KC)~\cite{Bear1972, Dullien1992, Koponen1997}, which relates permeability to porosity and specific surface and which we extend to a tensorial form whose off-diagonal components carry a sign, and a directional critical-path estimate (CP) based on the critical radius of Katz and Thompson~\cite{Katz1986}, the largest inscribed-sphere radius at which the pore space still percolates along $\hat d$. Their union ($38$ features) is denoted \emph{Phys}.

The union of Stats and Phys ($1857$ features) is the strongest non-topological descriptor set we consider and is denoted \emph{Baselines}; we use the same word for the whole class of non-topological descriptor sets (Stats, Phys and their components).

\subsection{Topological descriptors}
\label{sec:tda_new}

Both topological descriptors used in this work are based on persistent homology. Since topological data analysis (TDA) is not yet a standard tool in the physics of porous media, we first summarize the elements of persistent homology that are needed here, following its standard formulation~\cite{Edelsbrunner2002, Zomorodian2005}; comprehensive introductions can be found in~\cite{Edelsbrunner2010, Carlsson2009, Dlotko2024}.

\subsubsection{Persistent homology in a nutshell}

Topology describes a shape by quantities that do not change under continuous deformations: the number of separate pieces (connected components), the number of independent loops, i.e.\ tunnels that a closed string could thread without being able to contract, and the number of enclosed cavities. They are counted by the Betti numbers $\beta_0$, $\beta_1$ and $\beta_2$, which are cardinalates of the homology groups $H_0$, $H_1$ and $H_2$. For the pore space of a porous medium, $\beta_0$ counts isolated pore clusters, $\beta_1$ counts loops of the pore network, i.e.\ alternative flow paths that close on themselves around a piece of solid, and $\beta_2$ counts pockets of solid that are completely surrounded by pores (or, for the solid phase, pores that are completely surrounded by solid).

The Betti numbers of a single binary image say little about flow, since they do not contain any length scale. Persistent homology (PH)~\cite{Robins2016, Dlotko2024} adds the scale. Every voxel receives a number, the scalar value of a \emph{filtration function} $F$, and the structure is built up gradually: at a threshold $t$ we keep only the voxels with $F\le t$, the \emph{sublevel set} of $t$. As $t$ grows, voxels are added, and the topology of the sublevel set changes. A new component appears when a voxel is added that is not connected to anything present so far, and it disappears when it merges with an older component; a loop appears when a ring of voxels closes, and it disappears when the hole inside the ring is filled. Each topological class is therefore described by two numbers, the threshold at which it is \emph{born} and the threshold at which it \emph{dies}. The result is a list of (birth, death) pairs for every homology dimension, the \emph{persistence diagram}, often drawn as a \emph{barcode} in which every pair is a horizontal bar from its birth to its death. The difference between death and birth is the \emph{persistence} of the class: long bars correspond to large, robust features of the structure, short ones to small details.

A simpler summary, which does not pair births with deaths, is the Euler characteristic $\chi=\beta_0-\beta_1+\beta_2$. It can be computed without persistence, as the alternating count of the cells of the set, $\chi = n_\text{vertices}-n_\text{edges}+n_\text{faces}-n_\text{cubes}$, where every voxel is counted as a cube together with its faces, edges and vertices. Evaluated on the sublevel sets of $F$ as a function of the threshold $t$, it gives the Euler characteristic profile (ECP), $\chi(t)$. Because $\chi$ is cheap to compute, the ECP extends directly to several filtration functions $F_1,\dots,F_m$ (a multifiltration): $\chi(t_1,\dots,t_m)$ is the Euler characteristic of the set of voxels with $F_i\le t_i$ for all $i$.

The physical meaning of the diagram is determined entirely by the choice of $F$. The most common choice in porous media is the distance transform $\mathrm{dt}(\mathbf{x})$, the radius of the largest sphere centred at $\mathbf{x}$ that fits into the pore space~\cite{Robins2016, Herring2019, Moon2019}. If the pore space is grown from its widest parts to its narrowest, i.e.\ with the filtration $F=-\mathrm{dt}$, a pore body is born at its own inscribed radius, and two pore bodies merge, so that the younger of the two components dies, at the radius of the throat that connects them. A bar of $H_0$ thus records a pore size and a throat size. This is precisely the information on which classical estimates of permeability are based~\cite{Katz1986}.

\subsubsection{Why the filtration must depend on direction}

The distance transform, and therefore its persistence diagram, is invariant under rotations and reflections of the structure. A descriptor built on it can predict the scalar size of the pores and throats, but not a direction, and in particular it cannot predict the off-diagonal permeabilities, which change sign under a reflection (Section~\ref{sec:dataset}). Consider $k_{xy}$. The permeabilities measured along the two face diagonals of the $xy$ plane are
\[ k_{[110]} = \tfrac12(k_{xx}+k_{yy}) + k_{xy}, \qquad k_{[1\bar10]} = \tfrac12(k_{xx}+k_{yy}) - k_{xy}, \]
so that $k_{xy} = \tfrac12(k_{[110]}-k_{[1\bar10]})$ is half of the \emph{contrast} between the conductances along the two diagonals. A descriptor that is to predict $k_{xy}$ must therefore be able to tell the two diagonals apart, and the same holds for $k_{xz}$ with the pair $[101]$, $[10\bar1]$ and for $k_{yz}$ with $[011]$, $[01\bar1]$.

We therefore build the probing direction $\hat d$ into the filtration function, $F=F(\mathbf{x};\hat d)$, following our earlier work on the elastic stiffness tensor~\cite{Topolnicki2026TDA}. For every structure the filtration, and hence the persistence diagram, is computed separately for each of the nine directions listed at the beginning of this section. No direction is selected in advance: the features of all nine directions are concatenated into one feature vector, and the machine-learning model itself decides which directions are relevant for which component of the tensor. The geometric constructions are made directly along $\hat d$ on the periodic grid, so the binary image is never rotated or interpolated, all nine directions are treated identically, and the periodicity of the structure is preserved. We use two direction-aware filtrations, which differ in what they measure: the cone-based filtration measures the \emph{occupancy} of the neighbourhood of a solid voxel along $\hat d$, and the throat-scale filtration measures the \emph{width} of the pore channel along $\hat d$.

\begin{figure*}
\centering
\includegraphics[width=\linewidth]{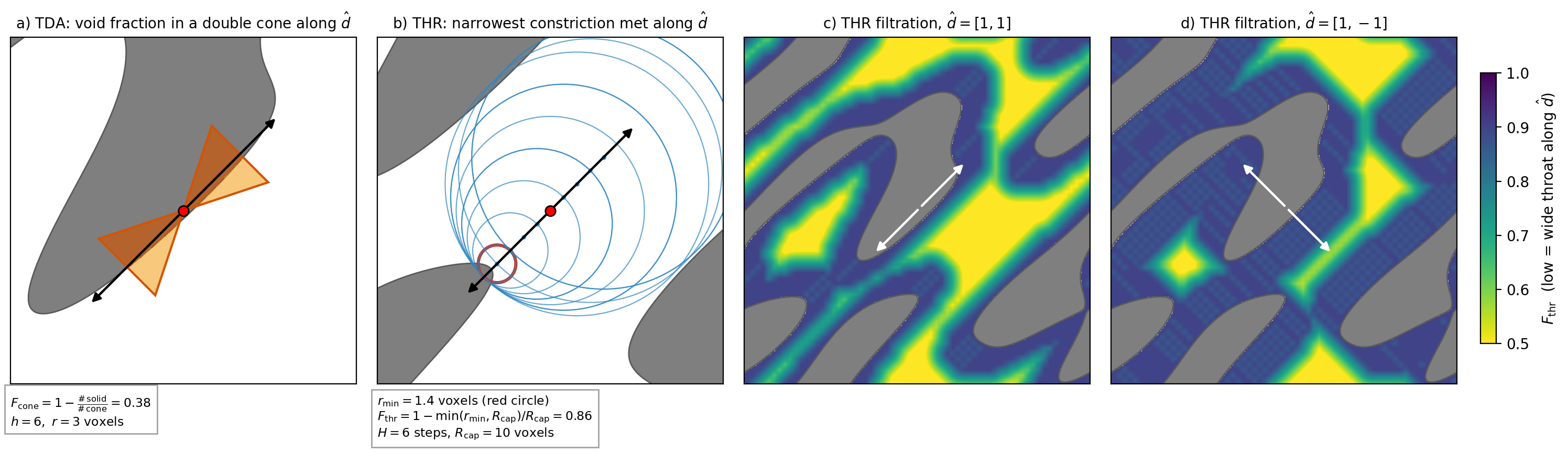}
\caption{The two direction-aware filtrations, illustrated on a periodic two-dimensional random-trigonometric structure (solid in grey); the actual filtrations are computed on three-dimensional grids with the same parameters. (a) TDA: for the solid voxel marked in red, the double cone of height $h=6$ and base radius $r=3$ voxels oriented along $\hat d=[1,1]$ (black arrow) is constructed; the filtration value is the void fraction inside the cone (light: void, dark: solid). (b) THR: from the pore voxel marked in red, a walk of at most $H=6$ steps is made along $\pm\hat d$; at every visited voxel the inscribed radius $\mathrm{dt}$ is shown as a blue circle, and the smallest one (red) is the narrowest constriction met along $\hat d$, which sets the filtration value. (c, d) The resulting THR filtration over the whole structure for $\hat d=[1,1]$ and $\hat d=[1,\bar1]$. Low values (yellow) mark channels that remain wide along $\hat d$; the two maps differ because the pore channels of the structure are elongated close to the $[1,1]$ direction.}
\label{fig:cartoon}
\end{figure*}

\subsubsection{Cone-based filtration (TDA)}

The cone-based filtration was introduced in~\cite{Topolnicki2026TDA}. It is defined on the solid phase. For every solid voxel we look along $\hat d$ through a double cone with its apex at the voxel (Figure~\ref{fig:cartoon}a) and count how much of the cone is void.

\begin{definition}
\label{def:cone_new}
A double cone with height $h$ and base radius $r$ attached at voxel $e$ and oriented along the unit vector $\hat d$ is the set of voxels $e+\mathbf{w}$ such that
\[ 0<|\mathbf{w}\cdot\hat d| \le h \quad\text{and}\quad \big|\mathbf{w}-(\mathbf{w}\cdot\hat d)\,\hat d\big| \le \frac{r\,|\mathbf{w}\cdot\hat d|}{h}, \]
with all indices taken modulo the grid size. The cone-based filtration of a solid voxel $e$ is the void fraction inside the cone,
\[ F_\text{cone}(e;\hat d) = 1-\frac{1}{|\mathrm{Cone}(e,r,h)|}\sum_{e'\in\mathrm{Cone}(e,r,h)} X(e'), \]
where $X=1$ for solid and $0$ for void voxels. Void voxels are assigned the value $1.25$, so that they enter the filtration after all solid voxels.
\end{definition}

A solid voxel inside a grain or a rod that is elongated along $\hat d$ sees mostly solid in its cone and enters early; a voxel of a thin plate crossed by $\hat d$, or one at a grain surface facing along $\hat d$, sees mostly void and enters late. The sublevel sets therefore grow the solid phase starting from its parts that are thick and continuous along $\hat d$, and the persistence diagram records how these parts connect into larger solid pieces ($H_0$), when they close loops of solid ($H_1$) and when they enclose pores ($H_2$).

The cone filtration is complemented, as in~\cite{Topolnicki2026TDA}, by two channels that describe the local orientation of the solid. For each solid voxel, the covariance matrix of the positions of the solid voxels within a sphere of radius $r_\text{PC}$ is computed, and the channels are $1-|\mathbf{v}_1\cdot\hat d|$ and $1-|\mathbf{v}_3\cdot\hat d|$, where $\mathbf{v}_1$ and $\mathbf{v}_3$ are the eigenvectors with the largest and the smallest eigenvalue. The first is small where the local solid is elongated along $\hat d$, the second where $\hat d$ crosses a plate. Unlike the channels $(1-|v_z|, 1-|v_y|)$ used in~\cite{Topolnicki2026TDA}, these two do not require a reference transverse axis, and the second one is needed for plate-like structures, whose long axis is not determined by the geometry. The three channels together define a three-parameter multifiltration of the solid phase, for which we compute the ECP $\chi(t_1,t_2,t_3)$ introduced above. The descriptor denoted \emph{TDA} consists of the persistence diagrams of $F_\text{cone}$ in dimensions 0, 1 and 2 and of the ECP. We use $h=6$, $r=3$ and $r_\text{PC}=4$ voxels.

\subsubsection{Throat-scale filtration (THR)}

The cone filtration measures the occupancy of a neighbourhood, a quantity related to the volume fractions and the second-order correlations of the structure. Critical-path analysis~\cite{Katz1986}, however, states that permeability is controlled by the narrowest constriction on the widest connected flow path, which is a \emph{radius}. The throat-scale filtration is designed to measure this radius along a given direction. It is defined on the pore space and combines the distance transform with a walk along $\hat d$ (Figure~\ref{fig:cartoon}b).

\begin{definition}
\label{def:thr_new}
For a pore voxel $\mathbf{x}$ and a unit direction $\hat d$, let $\mathbf{x}_s = \lfloor\mathbf{x} + s\hat d + \tfrac12\rfloor$ (modulo the grid size), $s=0,\pm1,\dots,\pm H$, be the voxels visited by a walk along $\pm\hat d$, where the walk in each sense stops at the first voxel that does not belong to the pore space. The throat-scale filtration is
\[ F_\text{thr}(\mathbf{x};\hat d) = 1-\frac{\min\left(r_{\min}(\mathbf{x};\hat d),\,R_\text{cap}\right)}{R_\text{cap}}, \qquad r_{\min}(\mathbf{x};\hat d) = \min_{s}\mathrm{dt}(\mathbf{x}_s), \]
and solid voxels are assigned the value $1.25$.
\end{definition}

In words, we stand at a pore voxel, walk at most $H$ steps forwards and backwards along $\hat d$, and record the narrowest inscribed radius $r_{\min}$ met on the way. A voxel in a channel that stays wide along $\hat d$ receives a large $r_{\min}$ and a low filtration value. A voxel from which the walk along $\hat d$ soon reaches a constriction, or the solid wall itself, where $\mathrm{dt}$ is about one voxel, receives a small $r_{\min}$ and a high filtration value, even if the voxel itself sits in the middle of a wide pore. The same voxel can therefore receive very different values for different directions: in a straight channel, it is ``wide'' when looked at along the channel and ``narrow'' when looked at across it (Figure~\ref{fig:cartoon}c,d). An obvious alternative, an anisotropic metric that is cheap along $\hat d$ and expensive across it, would read a channel that is \emph{blocked} along $\hat d$ as infinitely wide; taking the minimum along the walk avoids this. Because the filtration value is a radius, $r = R_\text{cap}(1-F_\text{thr})$, every birth and death of the persistence diagram can be read directly as a radius in voxels.

The sublevel sets of $F_\text{thr}$ grow the pore space from the channels that are widest along $\hat d$ to the narrowest ones, and the persistence diagram can be read in terms of pores and throats seen along $\hat d$. As a simple example, consider a channel along $\hat d$ made of two wide chambers with inscribed radius $5$ connected by a throat of radius $2$. The central voxels of each chamber, farther than $H$ steps from the throat, enter at $r\approx5$ as two separate components; the voxels whose walk reaches the throat enter only at $r=2$, and then the two chambers merge. The younger component dies, and its bar in $H_0$ runs from $r=5$ to $r=2$: it gives the chamber size and the throat size as seen along $\hat d$. If the same channel is probed across its axis, the walks of almost all voxels reach the solid wall within a few steps, all voxels enter late at small radii, and the bar is short. A class of $H_1$ is born when a ring of pore voxels closes around a region that has not yet entered, and it dies when that region is filled. If the region is a narrower part of the pore space, the death radius is the radius of this narrower passage; if it is solid, the class lives until the solid enters at the end of the filtration. Many short-lived $H_1$ and $H_2$ classes at small radii therefore indicate that, along $\hat d$, the pore space is broken up by many narrow constrictions.

The construction of THR was motivated by critical-path analysis~\cite{Katz1986, Ambegaokar1971, Hunt2001}, and the relation between the two can be stated precisely. In percolation theory, the conductance of a medium with a broad distribution of local conductances is controlled by the critical one, the smallest conductance that must still be included for a connected path to span the system; in a porous medium this is the critical radius $r_c$, the largest radius for which the set $\{\mathbf{x}:\mathrm{dt}(\mathbf{x})\ge r_c\}$ still percolates~\cite{Katz1986}. In the language of persistent homology, $r_c$ along $\hat d$ is a single event of the filtration of $-\mathrm{dt}$: the radius at which a cluster of the pore space first wraps around the periodic domain along $\hat d$. The persistence diagram records, in addition, every other merge of pore clusters, at radii larger and smaller than $r_c$, and THR makes these events depend on $\hat d$ through the walk. This additional information is what the off-diagonal components require. As shown in Section~\ref{sec:results}, the two diagonals of a pair have the same critical radius in $91\%$ of the structures, so that $r_c$ alone carries almost no information on the sign of the off-diagonal permeability, whereas THR, which records the whole sequence of throats seen along each diagonal, predicts it better than any other descriptor set.

Section~S9 of the Supplementary Information shows this on the two structures of Figure~\ref{fig:examples}, for the signed pair of diagonals $[011]$ and $[01\bar1]$ that determines $k_{yz}$. For the isotropic structure the two directions give practically identical barcodes, as they should, since its $k_{yz}$ is close to zero. For the strongly anisotropic structure, whose pore channels run close to $[011]$, the two barcodes differ strongly. Along the channels, the median radius of the pore voxels is $2.8$ voxels and the barcodes contain about $1500$ pairs in $H_1$ and $260$ in $H_2$. Across the channels, the median radius drops to $1.4$ voxels, and the pore space closes in many small loops and cavities, about $14800$ pairs in $H_1$ and $8400$ in $H_2$. This contrast between the two diagonals is the information from which the model reads the sign and the magnitude of $k_{yz}$.

The filtration has two parameters. The walk length $H$ sets the distance over which a constriction is sought; for $H=0$ the filtration reduces to the isotropic distance transform, which is the same for every direction. We use $H=6$ steps, about one pore diameter (the median inscribed radius of the pore voxels is $3.2$ voxels); this value was not tuned, and a sensitivity test for $H$ between 2 and 16 steps (Section~S11 of the Supplementary Information) shows that it gives the lowest mean absolute error of THR on all three off-diagonal components. The cap $R_\text{cap}=10$ voxels only bounds the filtration range; only $0.1\%$ of the pore voxels have a larger inscribed radius. $R_\text{cap}$ is a global constant rather than a per-structure maximum, because permeability scales as $r^2$ and the absolute size of the throats is the quantity of interest. The persistence diagrams of $F_\text{thr}$ in dimensions 0, 1 and 2 form the descriptor denoted \emph{THR}.

All persistence diagrams are computed with GUDHI~\cite{Maria2014}, with periodic boundary conditions in all three directions, the same as in the LBM simulations. In both descriptors, the persistence diagrams are converted into fixed-length feature vectors, persistence images, as described in Section~\ref{sec:training}. Classes that never die, i.e.\ that survive until the whole periodic domain has entered the filtration (one component, and the loops and cavities that wind around the periodic domain), give infinite persistence pairs, which are discarded.

\section{Model training}
\label{sec:training}

\paragraph{Vectorization.} Since the machine learning model requires vector-formatted inputs, persistence diagrams were converted into persistence images~\cite{JMLR:v18:16-337} with a resolution of $10\times10$, separately for homology dimensions 0, 1 and 2, and concatenated. Each persistence pair is weighted linearly by its persistence, and the image range and the Gaussian bandwidth are set separately for each homology dimension, with the bandwidth comparable to one pixel (Table~S8.1 in Section~S8 of the Supplementary Information); with the library defaults (constant weight and a bandwidth of one filtration unit) the images were found to be effectively of rank one, i.e.\ every diagram was painted as the same blob. Infinite (essential) persistence pairs are discarded. The Euler characteristic profiles were vectorized by sampling the three-parameter multifiltration on a $7\times7\times7$ grid. This gives, per direction, $300$ PH features and $343$ ECP features, and in total $2700$ PH and $3087$ ECP features for TDA and $2700$ PH features for THR.

\paragraph{Model.} Vectorized descriptors were used to train gradient-boosted decision trees with the CatBoost algorithm~\cite{Prokhorenkova2019}, with 1500 iterations, tree depth 6, learning rate 0.05, and the RMSE loss. The hyperparameters were fixed for all descriptor sets and all targets and were not tuned. No early stopping and no validation split were used; all models are trained for the full number of iterations. A separate model is trained for each descriptor set and each of the six tensor components (single-target regression); the targets are the raw tensor components in lattice units. The fixed number of iterations lies on the plateau of the test error for all descriptor sets; longer training and early stopping change $R^2$ by at most $0.002$ and $0.014$, respectively, and leave all conclusions unchanged (Section~S10 of the Supplementary Information).

\paragraph{Cross-validation.} Model training was performed using $k$-fold cross-validation with $k=5$. The dataset was shuffled and partitioned into five folds of 1000 structures; for each of the five models, one fold was held out as the test set and the remaining 4000 structures were used for training. Each structure contributes a single row containing the descriptors for all nine directions together with all six tensor components, so a given structure's descriptors and tensor components are always assigned jointly to the same fold, and data leakage across directions or tensor components is avoided by construction. All metrics reported in this paper were computed exclusively on the test folds, and each structure served as a test instance exactly once. All descriptor sets are evaluated on identical folds, which allows a paired comparison.

\paragraph{Metrics.} We report the coefficient of determination $R^2$ and the mean absolute error (MAE) of the predicted tensor components with respect to those computed with the LBM, Eq.~\eqref{eq:Kij} , both computed on each test fold and averaged over the five folds. Because the folds are shared, two descriptor sets $A$ and $B$ can be compared fold by fold: we report the mean difference $\Delta R^2 = \langle R^2_A - R^2_B\rangle$, the mean relative change of the MAE, $\langle \mathrm{MAE}_A/\mathrm{MAE}_B - 1\rangle$, and the number of folds in which $A$ has the lower MAE.
The paired comparisons for all relevant pairs of descriptor sets are given in Section~S3 of the Supplementary Information.

\section{Results}
\label{sec:results}

\begin{figure*}
\centering
\includegraphics[width=\linewidth]{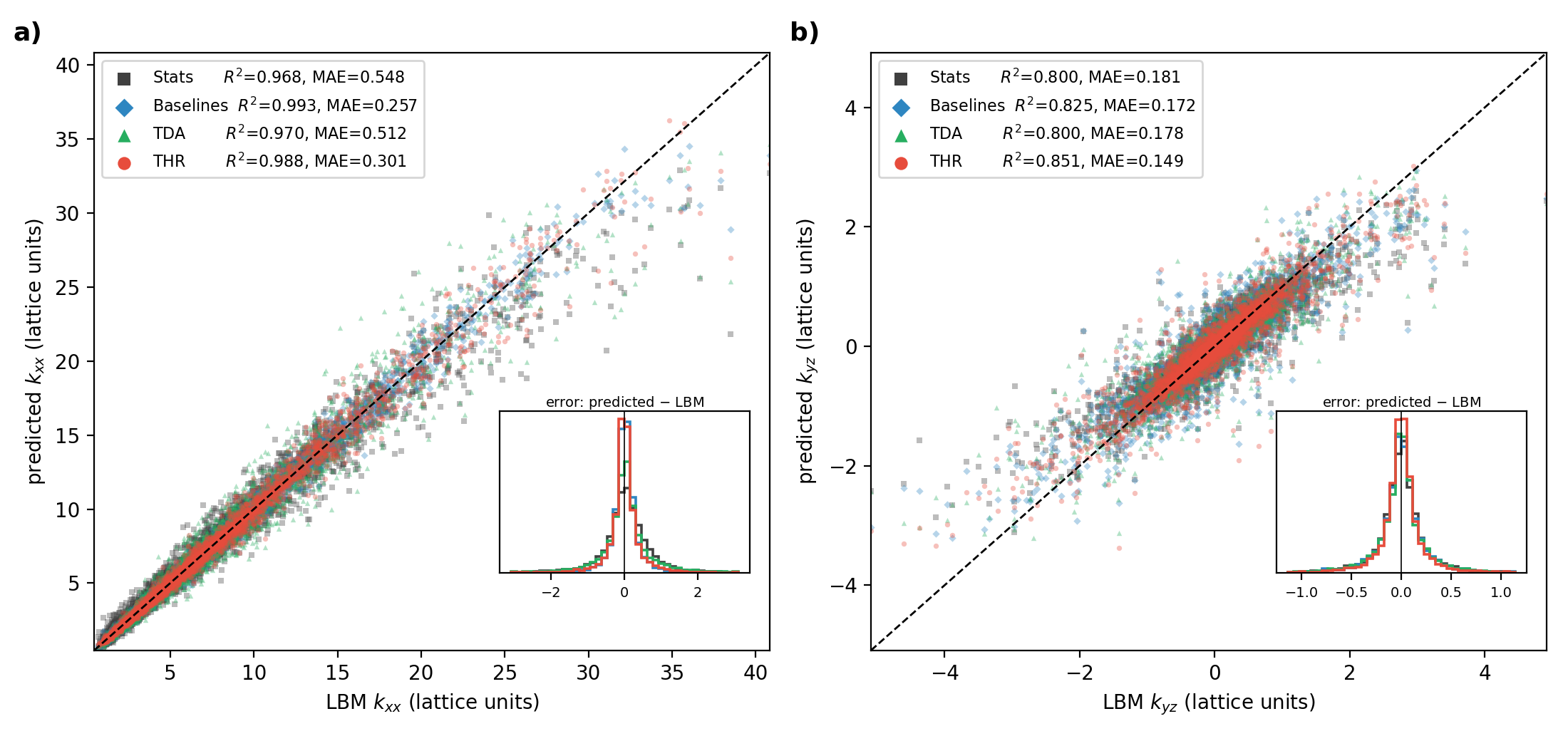}
\caption{Predicted vs.\ LBM-computed permeability for (a) the diagonal component $k_{xx}$ and (b) the off-diagonal component $k_{yz}$, comparing THR (red circles), TDA (green triangles), Baselines (blue diamonds) and Stats (grey squares). Each panel pools the test-set predictions of all 5 cross-validation folds; the dashed line marks perfect prediction. $R^2$ and MAE are averaged across the 5 folds. The inset shows the distribution of the prediction error, predicted $-$ LBM. Lattice units.}
\label{fig:scatter}
\end{figure*}

\begin{figure*}
\centering
\includegraphics[width=\linewidth]{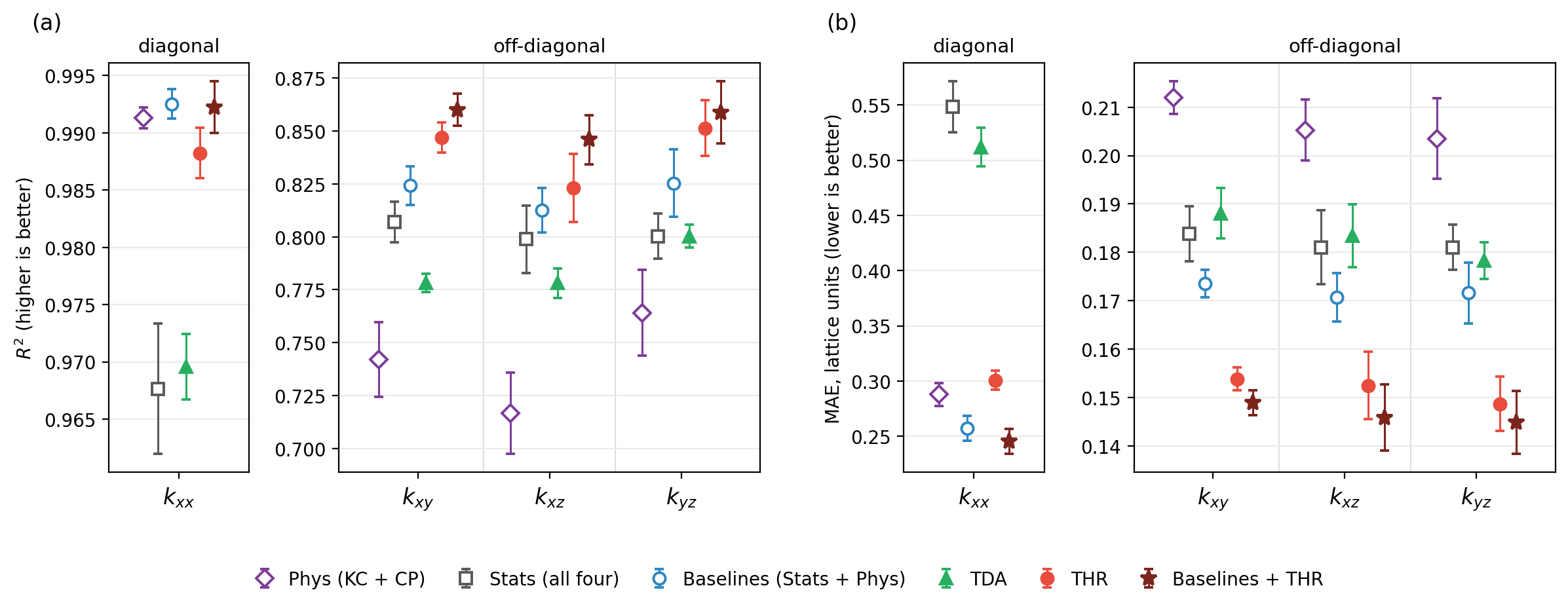}
\caption{Accuracy of the main descriptor sets for the diagonal component $k_{xx}$ and the three off-diagonal components: (a) the coefficient of determination $R^2$ and (b) the mean absolute error (MAE) of the predicted components with respect to the LBM-computed ones, in lattice units. Markers show the mean over the five test folds and error bars the standard deviation across the folds. Open markers: non-topological descriptor sets; filled markers: sets that contain topological descriptors. In each panel $k_{xx}$ has its own vertical axis, because its scale differs from that of the off-diagonal components. The values for all descriptor sets and all six components are given in Tables~S4.1 and~S4.2 of the Supplementary Information.}
\label{fig:summary}
\end{figure*}

Figure~\ref{fig:scatter} shows the predicted-versus-LBM scatter plots for one diagonal ($k_{xx}$) and one off-diagonal ($k_{yz}$) component, and Figure~\ref{fig:summary} summarizes $R^2$ and the MAE of the main descriptor sets for $k_{xx}$ and the three off-diagonal components. The diagonal components $k_{yy}$ and $k_{zz}$ behave like $k_{xx}$. The values of $R^2$ and MAE for all descriptor sets and all six components are given in Tables~S4.1 and~S4.2 of the Supplementary Information, and the fold-by-fold paired comparisons, from which the relative changes of the MAE and the numbers of folds won quoted below are taken, in Table~S3.1.

The diagonal components are predicted with high accuracy by almost all methods, with $R^2>0.96$ for every competitive descriptor set. 

On $k_{xx}$ the best single non-topological set is Baselines ($R^2=0.993$), and the Kozeny--Carman estimate alone already reaches $R^2=0.987$. THR is close behind ($R^2=0.988$), but its MAE is $17\%$ higher than that of Baselines. Adding THR to Baselines leaves $R^2$ unchanged and lowers the MAE by about $5\%$. On the diagonal, the classical physical estimate is therefore sufficient, and topology adds little to it. This is also visible in Figure~\ref{fig:scatter}a, where the point clouds of all four methods collapse onto the identity line and differ only in the spread at the largest permeabilities.

The picture changes for the off-diagonal components. Here a single topological descriptor, THR, predicts the off-diagonal permeabilities more accurately than every non-topological descriptor set, including the union of all four statistical descriptors (Stats), the physics-based estimators (Phys) and the union of both (Baselines), in both $R^2$ and MAE and in every fold (Figure~\ref{fig:summary}). The Kozeny--Carman estimate, the classical physical predictor of permeability, has a higher MAE than THR on all six components of the tensor, the diagonal ones included. Here THR is the best single descriptor set on all three components ($R^2=0.847$, $0.823$ and $0.851$ for $k_{xy}$, $k_{xz}$ and $k_{yz}$), ahead of Baselines ($0.824$, $0.812$, $0.825$), Stats ($0.807$, $0.799$, $0.800$) and TDA ($0.778$, $0.778$, $0.800$). THR also has the lowest MAE of all single sets, $16$--$18\%$ lower than Stats and $11$--$13\%$ lower than Baselines, and it is more accurate than both in every fold. In Figure~\ref{fig:scatter}b the THR predictions form the tightest cloud around the identity line, and its error distribution is the narrowest. All four methods underestimate the largest values of $|k_{yz}|$ (above $\approx2$ lattice units), i.e.\ they pull the tails of the distribution towards zero.

THR also contains information that is missing from the non-topological descriptors. Adding it to Baselines raises $R^2$ by $0.033$--$0.036$ and lowers the MAE by $14$--$16\%$ on every off-diagonal component, in every fold. Baselines + THR is the best descriptor set overall ($R^2=0.860$, $0.846$ and $0.859$), and Stats + THR is practically equal to it. The same holds when THR is added to the physics-based estimators alone: Phys + THR ($R^2=0.853$, $0.839$ and $0.859$) is as accurate as Baselines + THR, and compared with Phys it raises $R^2$ by $0.09$--$0.12$ and lowers the MAE by $29$--$30\%$.

The two topological descriptors differ in the filtration function, and this difference matters. TDA, built on the occupancy of a cone, is weaker than THR on every component, and on $k_{xy}$ and $k_{xz}$ it does not even reach the accuracy of Stats. Added to Baselines, it still lowers the MAE by $5$--$8\%$ on the off-diagonal components, about half of the gain obtained with THR. The same holds for Phys: adding TDA to it raises $R^2$ by $0.06$--$0.09$ and lowers the MAE by $17$--$18\%$ on the off-diagonal components, in every fold, against $0.09$--$0.12$ and $29$--$30\%$ for THR, and Phys + TDA ($R^2=0.804$, $0.805$ and $0.822$) remains less accurate than THR alone. We attribute this to what the two filtrations measure: the occupancy of a cone is to a large extent a re-encoding of the second-order statistics already contained in the two-point correlation function, while the throat radius is the quantity that critical-path analysis identifies as controlling permeability~\cite{Katz1986}.

The off-diagonal components also illustrate the role of symmetry. A reflection of the structure across a coordinate plane changes the sign of the off-diagonal components that involve the reflected axis (e.g.\ $x\to-x$ flips $k_{xy}$ and $k_{xz}$), and it exchanges the two face diagonals of the corresponding pair of directions (e.g.\ $[110]$ and $[1\bar10]$). A scalar descriptor is unchanged by the reflection and cannot predict these components, whereas a directional descriptor evaluated on both diagonals can carry the sign in the contrast between them. Two of the directional descriptors nevertheless remain close to zero on the off-diagonal components, although both are informative on the diagonal. The porosity profile reaches only $R^2=0.04$--$0.05$ ($0.78$ on $k_{xx}$). Its features are the solid fractions of the individual lattice planes normal to $\hat d$, indexed by their position along $\hat d$; a translation of the periodic structure, which leaves the permeability unchanged, shifts them cyclically, so that for structures with random phases the position-resolved features carry little information beyond their mean. The directional critical radius CP falls below zero ($R^2\approx-0.10$, against $0.96$ on $k_{xx}$). A cluster of the pore space that percolates along two coordinate axes also percolates along both diagonals of their plane, so the two diagonals of a pair have the same critical radius in $91\%$ of the structures, and CP carries almost no information on the sign. The Kozeny--Carman tensor, which inherits a sign from the fabric tensor, reaches $R^2=0.73$--$0.75$, and the two-point correlation function alone $0.77$--$0.79$; the latter is the component of Stats that carries most of its accuracy.

Finally, we checked that these results do not depend on the particular random partition of the data into folds. The whole comparison was repeated for two additional seed values, which reshuffle the folds and change the random state of CatBoost (Section~S3 of the Supplementary Information). The standard deviation of $R^2$ between repetitions is below $0.01$ for every descriptor set and component, several times smaller than the differences discussed above, and all conclusions are unchanged. In particular, Baselines + THR is more accurate than Baselines on every off-diagonal component in all 15 folds of the three repetitions, with $\Delta R^2=0.034$--$0.035$ and a $15$--$16\%$ lower MAE on average. Phys + TDA is more accurate than Phys on every off-diagonal component in all 15 folds as well ($\Delta R^2=0.055$--$0.099$, $17$--$19\%$ lower MAE), but its gain remains smaller than that of Phys + THR ($0.097$--$0.130$, $29$--$30\%$).

\begin{figure*}
\centering
\includegraphics[width=\linewidth]{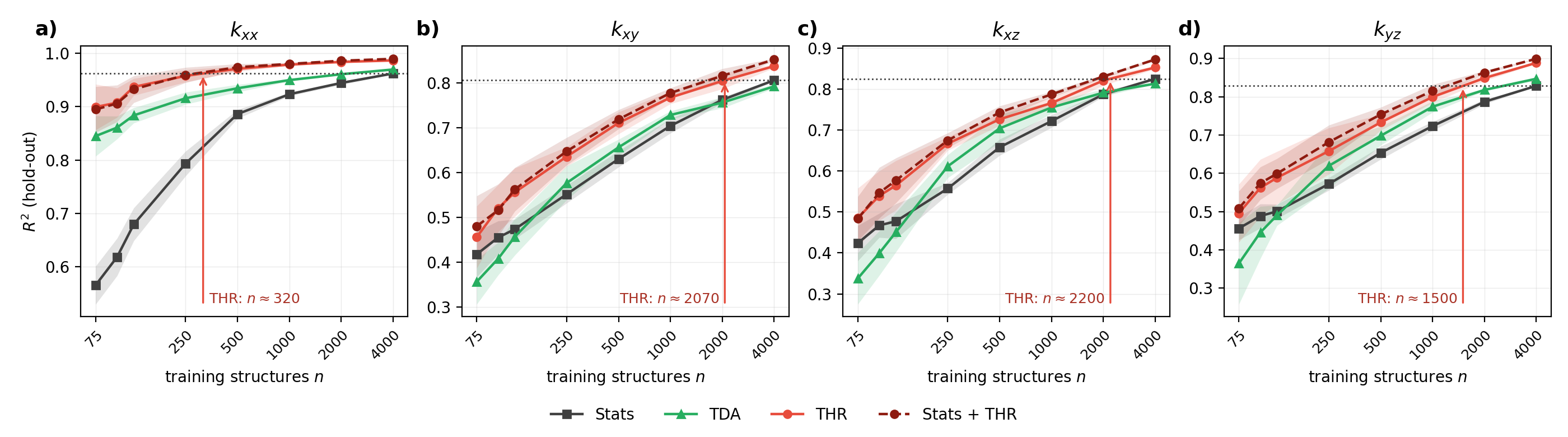}
\caption{Accuracy as a function of the number of training structures $n$ for (a) the diagonal component $k_{xx}$ and (b--d) the three off-diagonal components: $R^2$ on a fixed hold-out set (one fifth of the dataset) for Stats (grey squares), TDA (green triangles), THR (red circles) and Stats + THR (dark red, dashed). Lines show the mean and shaded bands the standard deviation over five random training subsets of each size. The dotted line marks $R^2$ of Stats trained on $n=4000$ structures, and the arrow the number of structures with which THR reaches the same accuracy.}
\label{fig:learning_curve}
\end{figure*}

We also examined how the accuracy depends on the amount of training data (Figure~\ref{fig:learning_curve}). For this test, one fifth of the structures were set aside as a balanced, fixed hold-out set and never used for training. Training sets of $n=75$ to $4000$ structures were drawn from the remaining ones, five times for each $n$, and every model was evaluated on the same hold-out set, so that the curves of different descriptor sets can be compared point by point. For a given random draw, every smaller training set is contained in every larger one.

THR is more accurate than Stats at every training-set size and on all four components. On the off-diagonal components it raises $R^2$ by $0.03$--$0.11$ and lowers the MAE by $5$--$22\%$, and from $n=250$ on it is more accurate in each of the five training subsets. Put differently, THR reaches the accuracy that Stats attain with 4000 training structures with roughly half of the data, about $1500$--$2200$ structures. On $k_{xx}$ the advantage is largest for small training sets ($R^2=0.90$ against $0.57$ for Stats at $n=75$). TDA shows no such consistent advantage: on the off-diagonal components it is less accurate than Stats for the smallest training sets ($n\le125$).

\section{Summary}
\label{sec:summary}
We asked whether direction-aware topological descriptors can predict the full permeability tensor of three-dimensional porous structures from their geometry. To answer it, we generated 5000 periodic random-trigonometric structures with controlled anisotropy in a random principal frame, computed their permeability tensors with the lattice Boltzmann method, and compared two topological descriptors with statistical microstructure descriptors and with physics-based estimators, in paired, fold-by-fold comparisons.

The diagonal and the off-diagonal components behave differently. The diagonal components are predicted with $R^2\approx0.99$ once the physics-based estimators are included (Baselines), and the Kozeny--Carman estimate alone reaches $R^2=0.987$. Topology adds little to them. The off-diagonal components are small and change sign under reflections, so scalar descriptors cannot predict them, and directional descriptors can predict them only through the contrast between the two diagonals of a pair. The porosity profile, whose position-resolved features are shifted by any translation of the structure, reaches only $R^2\approx0.05$ on them, and the critical radius, which is the same for both diagonals of a pair in $91\%$ of the structures, falls below zero. They are the part of the tensor where the choice of descriptor matters.

The main result concerns the throat-scale filtration THR, built on the distance transform of the pore space along a given direction. On the off-diagonal components it is the most accurate single descriptor set ($R^2=0.82$--$0.85$) : a single topological descriptor predicts them better than all statistical microstructure descriptors and physics-based estimators taken together, and it has a lower mean absolute error than the Kozeny--Carman estimate on every component of the tensor, and its mean absolute error is $11$--$13\%$ lower than that of the strongest non-topological union, Baselines, in every fold. It also carries information that the non-topological descriptors lack: adding THR to Baselines lowers the mean absolute error by $14$--$16\%$ on every off-diagonal component, in all 15 folds of three repeated cross-validations. The cone-based filtration TDA of our earlier work on the elastic stiffness tensor gives about half of this gain. We attribute the difference to what the two filtrations measure. The occupancy of a cone largely repeats the second-order statistics already contained in the two-point correlation function, while the throat radius is the quantity that critical-path analysis identifies as controlling permeability.

THR is also more data-efficient than the statistical descriptors. It is more accurate than Stats at every training-set size from 75 to 4000 structures, and it reaches the accuracy that Stats attain with 4000 training structures with about 1500--2200 structures on the off-diagonal components.

Two further tests, reported in the Supplementary Information, support these conclusions. A single model that predicts the whole tensor at once is less accurate than separate models for every component, in all 180 comparisons (Section~S6). On sub-volumes of Berea sandstone with an imposed affine stretch (Section~S7), the off-diagonal components are dominated by the stretch and the second-order descriptors capture most of them, so the gain from topology is smaller. It still persists: Baselines + TDA is the best set on $k_{xy}$ and $k_{yz}$, with a $5$--$7\%$ lower mean absolute error than Baselines, reproduced in three repetitions of the cross-validation.

These results have limitations. The anisotropy of all structures studied here is imposed, by a map in Fourier space for the synthetic structures and by an affine stretch for Berea, and all samples have the same size of $80^3$ voxels. Whether the gain from THR carries over to naturally anisotropic rocks, such as layered or sheared sandstones, remains to be tested. Within these limits, the direction-aware throat-scale filtration outperforms all non-topological descriptors on the off-diagonal permeabilities, and it does so for a physically interpretable reason.

The next step of our work will be the tortuosity tensor. It describes the geometry of the flow paths rather than the average flux and depends on the whole velocity field, which makes it a further step in complexity beyond the permeability tensor studied here.

\subsection*{Acknowledgements}
The calculations were made with the support of the Interdisciplinary Center for Mathematical and Computational Modeling of the University of Warsaw (ICM UW) under the computational grant no.\ g107-2962. The computations in this work were partially supported by the Center for Artificial Intelligence at Adam Mickiewicz University. We would like to thank Prof.\ K.~Jassem and Dr.\ B.~Naskr\k{e}cki for granting us access to the Center's infrastructure.

\subsection*{Funding}
R.~Topolnicki and P.~D{\l}otko acknowledge the support of the Dioscuri program initiated by the Max Planck Society, jointly managed with the National Science Centre (Poland), and mutually funded by the Polish Ministry of Science and Higher Education and the German Federal Ministry of Education and Research. 
M. Matyka acknowledges the financial support from the Slovenian Research And Innovation Agency (ARIS) research core funding No.\ P2-0095.
Funded by the National Science Centre, Poland, under the OPUS call in the Weave programme 2021/43/I/ST3/00228. This research was funded in whole or in part by the National Science Centre (2021/43/I/ST3/00228).

\section*{Declarations}
\subsection*{Author Contributions Statement}
R.T.\ conceived the study, generated the porous structures, performed the lattice Boltzmann simulations, conceived the throat-scale (THR) descriptor, implemented the methods, prepared and trained the machine-learning models, and wrote the manuscript. P.D.\ provided consultation on topological data analysis, took part in discussions and helped to prepare the manuscript. M.M.\ provided consultation on fluid flow in porous media, took part in discussions and helped to prepare the manuscript.
\subsection*{Competing Interests}
The authors declare no competing interests.

\bibliographystyle{sn-mathphys-num}
\bibliography{refs}

\includepdf[pages=-,pagecommand={\thispagestyle{empty}}]{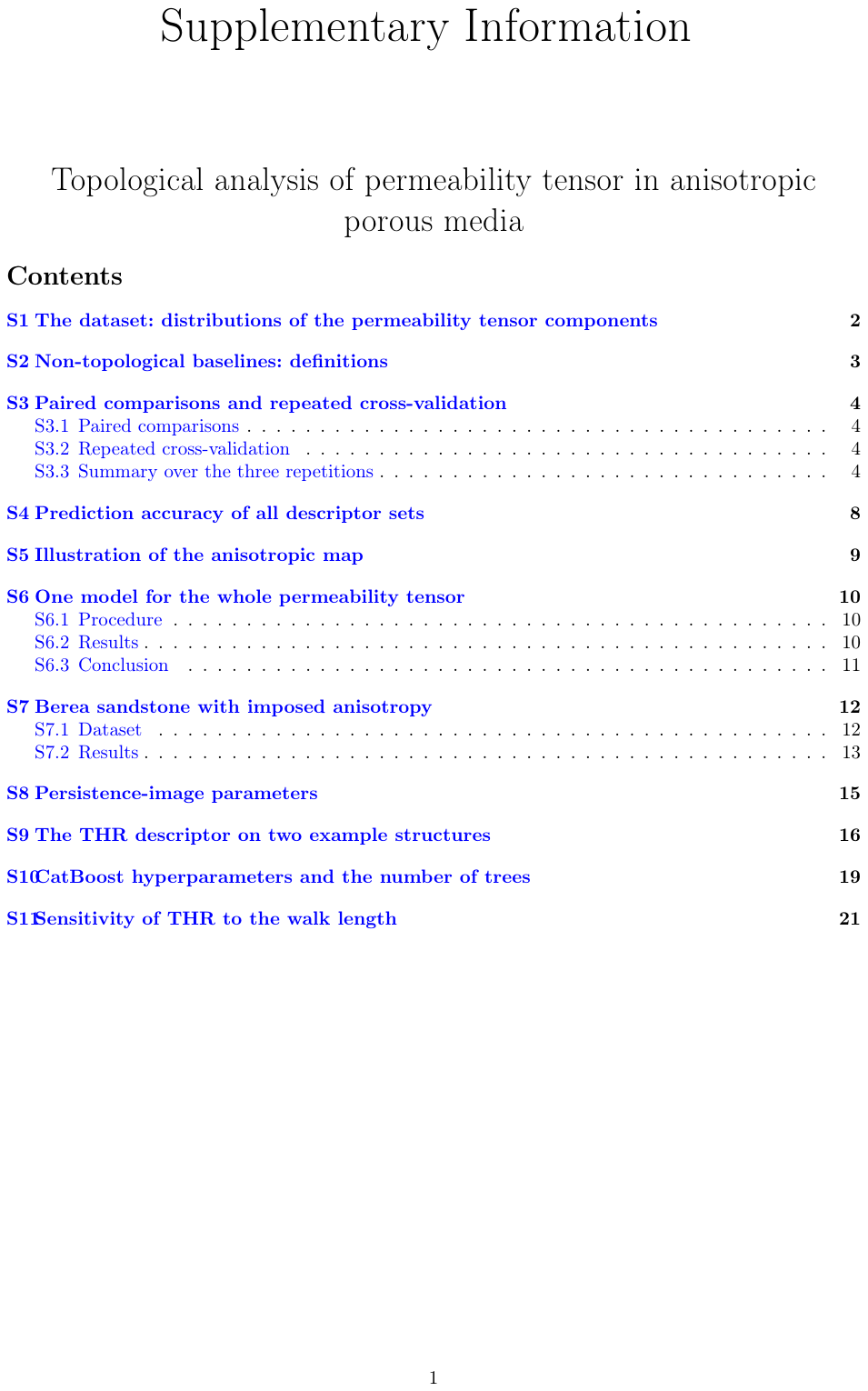}

\end{document}